\documentclass[conference]{IEEEtran}
\IEEEoverridecommandlockouts
\usepackage{cite}
\usepackage{amsmath,amssymb,amsfonts}
\usepackage{algorithmic}
\usepackage{graphicx}
\usepackage{textcomp}
\usepackage{xcolor}
\usepackage{subcaption}
\usepackage{multirow}
\usepackage{url}
\usepackage{makecell} 
\usepackage{booktabs} 
\usepackage{etoolbox} 
\usepackage{caption}   

\def\BibTeX{{\rm B\kern-.05em{\sc i\kern-.025em b}\kern-.08em
    T\kern-.1667em\lower.7ex\hbox{E}\kern-.125emX}}

\newcommand{\insertfig}{
    \setcounter{figure}{0}
    \captionsetup{type=figure}
    \centering
    \begin{subfigure}[t]{0.33\textwidth}
        \centering
        \includegraphics[width=\linewidth]{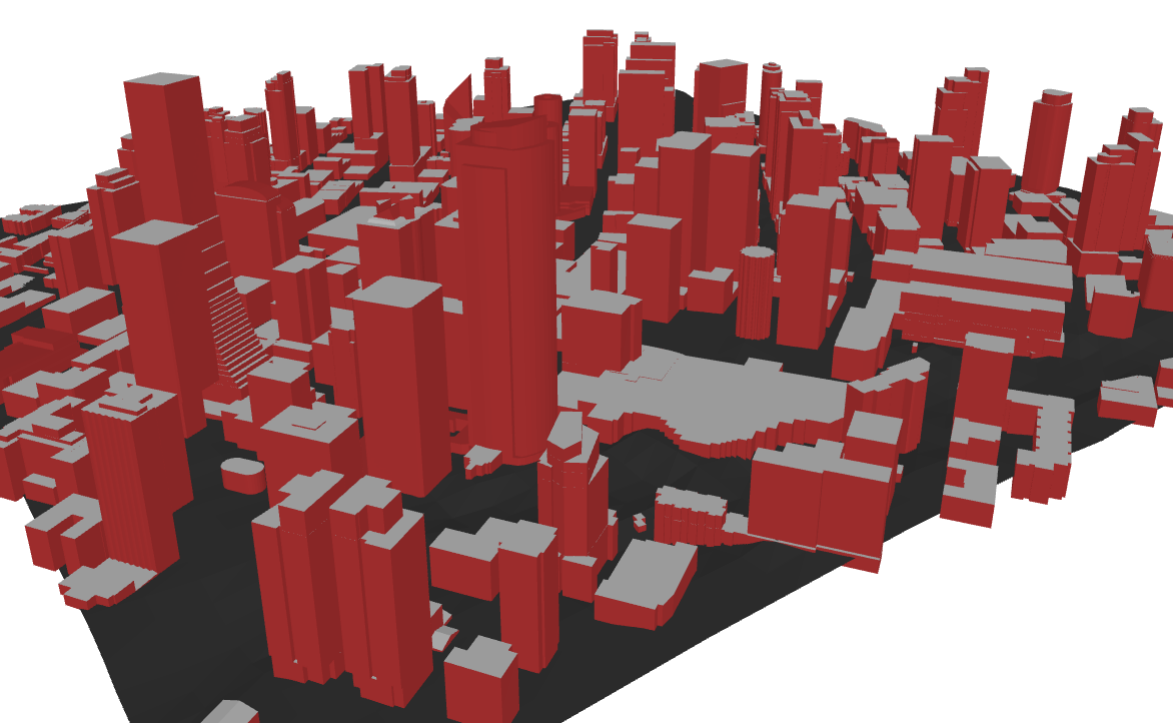}
        \caption{Seattle}
        \label{fig:seattle}
    \end{subfigure}%
    \hfill
    \begin{subfigure}[t]{0.33\textwidth}
        \centering
        \includegraphics[width=\linewidth]{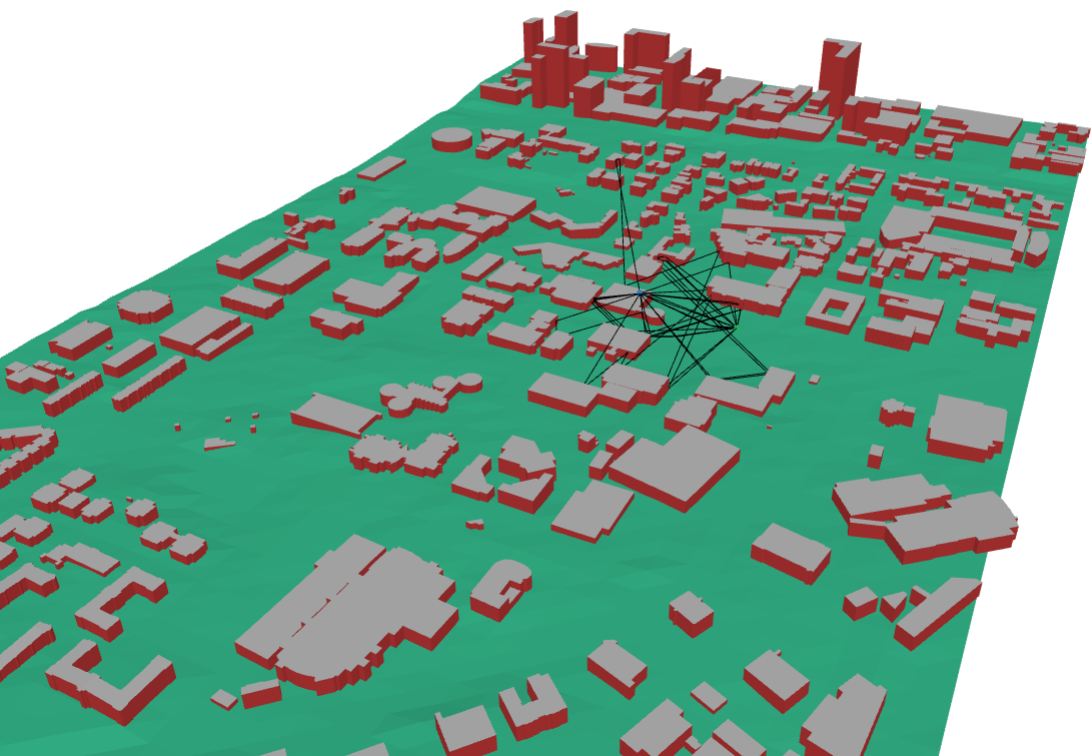}
        \caption{Georgia Tech - Atlanta}
        \label{fig:georgia_tech}
    \end{subfigure}%
    \hfill
    \begin{subfigure}[t]{0.33\textwidth}
        \centering
        \includegraphics[width=\linewidth]{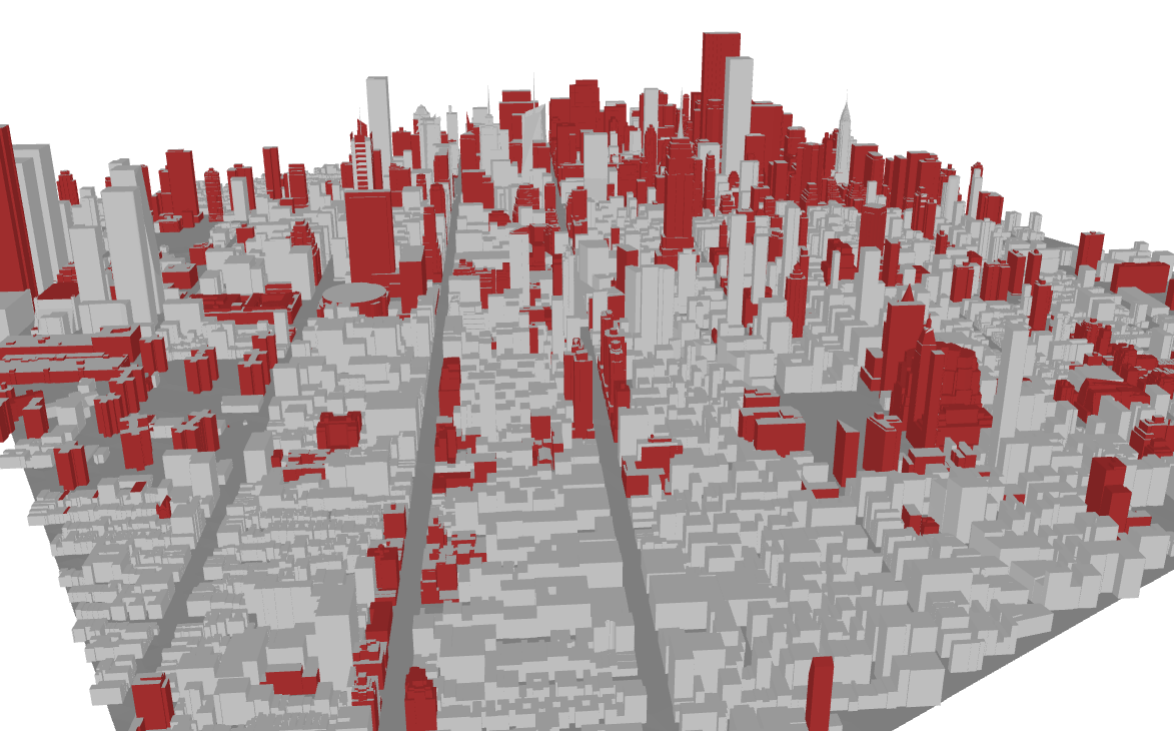}
        \caption{Manhattan}
        \label{fig:manhattan}
    \end{subfigure}
    \caption{Seattle, Georgia Tech, and Manhattan represented in Sionna RT through automatic geometry extraction pipeline.}
    \label{fig:scenes_combined}
}

\makeatletter
\apptocmd{\@maketitle}{\insertfig}{}{}%
\makeatother

\title{OpenGERT: Open Source Automated Geometry Extraction with Geometric and Electromagnetic Sensitivity Analyses for Ray-Tracing Propagation Models 
\thanks{This work was supported by Polaris Wireless. (Corresponding author: Serhat Tadik.)}
}

\author{
    Serhat Tadik${}^1$, Rajib Bhattacharjea${}^{1,*}$, Johnathan Corgan${}^{2,*}$,\\
    David Johnson${}^3$, Jacobus Van der Merwe${}^3$, Gregory D. Durgin${}^1$\\
    \footnotesize%
    \begin{tabular}{ccc}
        ${}^1$Georgia Institute of Technology & ${}^2$Corgan Labs & ${}^3$University of Utah \\
    \end{tabular}\\%
    \footnotesize ${}^*$Equal contribution%
}

\begin{document}

\maketitle

\begin{abstract}
Accurate RF propagation modeling in urban environments is crucial for developing high-fidelity digital spectrum twins and optimizing wireless communication systems. This paper introduces \textit{OpenGERT}, an open-source automated \underline{G}eometry \underline{E}xtraction tool for \underline{R}ay \underline{T}racing. \textit{OpenGERT} automates the collection and processing of terrain and building data from multiple sources, including OpenStreetMap, Microsoft Global ML Building Footprints, and USGS terrain elevation data. Leveraging the Blender Python API, the tool creates detailed urban models necessary for high-fidelity ray-tracing simulations specifically designed for NVIDIA Sionna RT.
Moreover, we conduct sensitivity analyses to assess the impact of variations in building height, position, and electromagnetic material properties on the accuracy of ray-tracing models. Specifically, we present pairwise dispersion plots of channel statistics—such as path gain, mean excess delay, delay spread, link outage, and Rician \textit{K}-factor—in response to perturbations to analyze the covariance of different channel statistics. The analyses also explore how the sensitivities of these statistics change as a function of distance from the transmitters. Additionally, we provide visualizations of the variance of the channel statistics within the scene for selected transmitter locations to offer deeper insights. Our study reports results from the Munich and Etoile scenes, each featuring 10 transmitter locations. For each transmitter location, we apply perturbations across five different types, 50 perturbations for each: material, position, height, height and position combined, and all combined. The findings reveal that, assuming the initial material properties of buildings are roughly accurate, minor perturbations in permittivity and conductivity do not significantly alter the channel statistics. In contrast, variations in building height and position have a considerable impact on all the statistics, even with a noise standard deviation of 1 meter for building heights and 0.4 meters for building positions. These results highlight the importance of precise environmental modeling in achieving reliable propagation predictions, which are essential for the deployment of digital spectrum twins and advanced communication networks. Finally, we share the code for geometry extraction and sensitivity analyses at \url{https://github.com/serhatadik/OpenGERT/} to facilitate further experimentation and development.

\end{abstract}


\section{Introduction}

\begin{figure*}[htbp]
\centering
\includegraphics[width=0.95\linewidth]{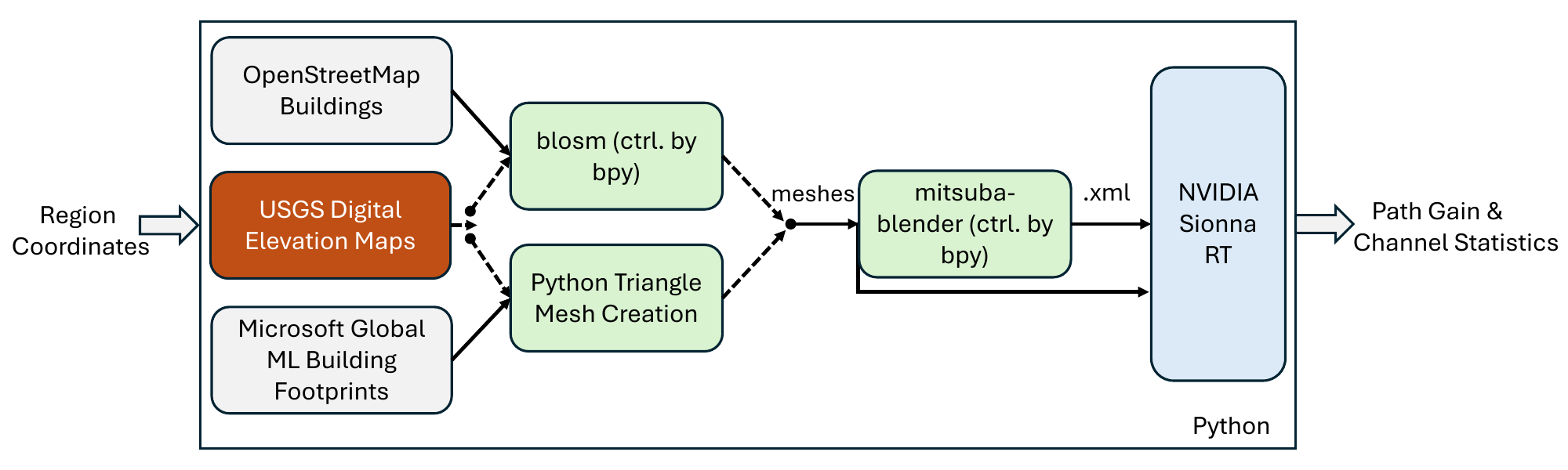}
\caption{Geometry extraction pipeline.}
\label{fig:ge_pipeline}
\end{figure*}

\subsection{Digital Spectrum Twins}

The concept of a digital twin can potentially transform many industries by providing dynamic, digital, and cloud-based representations of physical assets, systems, and environments. These digital counterparts enable real-time monitoring, simulation, and predictive analysis improving decision-making and operational efficiency. In the context of wireless communications, this concept has evolved into \textit{digital spectrum twins} (DST), which are high-fidelity digital replicas of the RF spectrum that keep track of the RF activity in a region through spectrum measurements and propagation modeling \cite{DST, DST2, DST3} with applications in 5G, and 6G networks \cite{6g_dt_nets, alloc_dt, dt_edge_network, khan2021digitaltwinenabled, dt_5g,  bostontwin, colosseum_dt}, and multi-media streaming \cite{mm_stream_dt1, mm_stream_dt2}. DSTs are envisioned to play an important role in designing, testing, and optimizing wireless networks, especially as the demand for low-latency, and high data-rate connectivity and the complexity of communication systems continue to grow. They can also be important enablers of various spectrum-sharing and access technologies \cite{powder-rdz}.

A fundamental challenge in developing accurate DSTs lies in the detailed, accurate extraction and modeling of environmental geometries, such as buildings and terrain. Traditionally, this geometry extraction process has been manual, labor-intensive and susceptible to large-scale inaccuracies, limiting the scalability and adaptability of propagation models across different regions and scenarios. Automating this process is fundamental for accelerating the development and large-scale adoption of DSTs.

Ray-tracing models are considered one of the best candidates for use in DSTs due to their accuracy in simulating electromagnetic wave propagation through interactions with environmental features such as reflection, diffraction, and scattering. However, the extent to which the accuracy of these models relies on the precision of the environmental geometry data and the electromagnetic properties of building and terrain materials remains uncertain.

The contributions of this paper can be summarized as:
\begin{itemize}
    \item An open source implementation of a method for automating the geometry extraction process for ray-tracing propagation models specifically for the NVIDIA Sionna Ray-Tracing tool \cite{sionnart} using multiple data sources including OpenStreetMap \cite{openstreetmap}, and Microsoft Global ML Building Footprints \cite{globalmlbuildingfootprints} dataset for building data and U.S. Geological Survey (USGS) \cite{usgs} for terrain data.
    \item Open source implementation of building geometry and material perturbation pipeline designed to introduce Gaussian noise to building heights, positions, relative permittivity, and conductivity, enabling the community to explore the effects of uncertainty in these parameters.
    \item An extensive sensitivity analysis of the ray-tracing model to perturbations in building heights, positions, and material properties. Specifically, we analyze the impact on key channel statistics such as path gain, mean excess delay, delay spread, and link reliability. Additionally, we examine how these statistics co-vary under identical perturbations and how they vary with different transmitter-receiver separation distances.
\end{itemize}

\begin{figure*}[htbp]
    \centering
    \begin{subfigure}{0.4\textwidth}
        \centering
        \includegraphics[width=\linewidth]{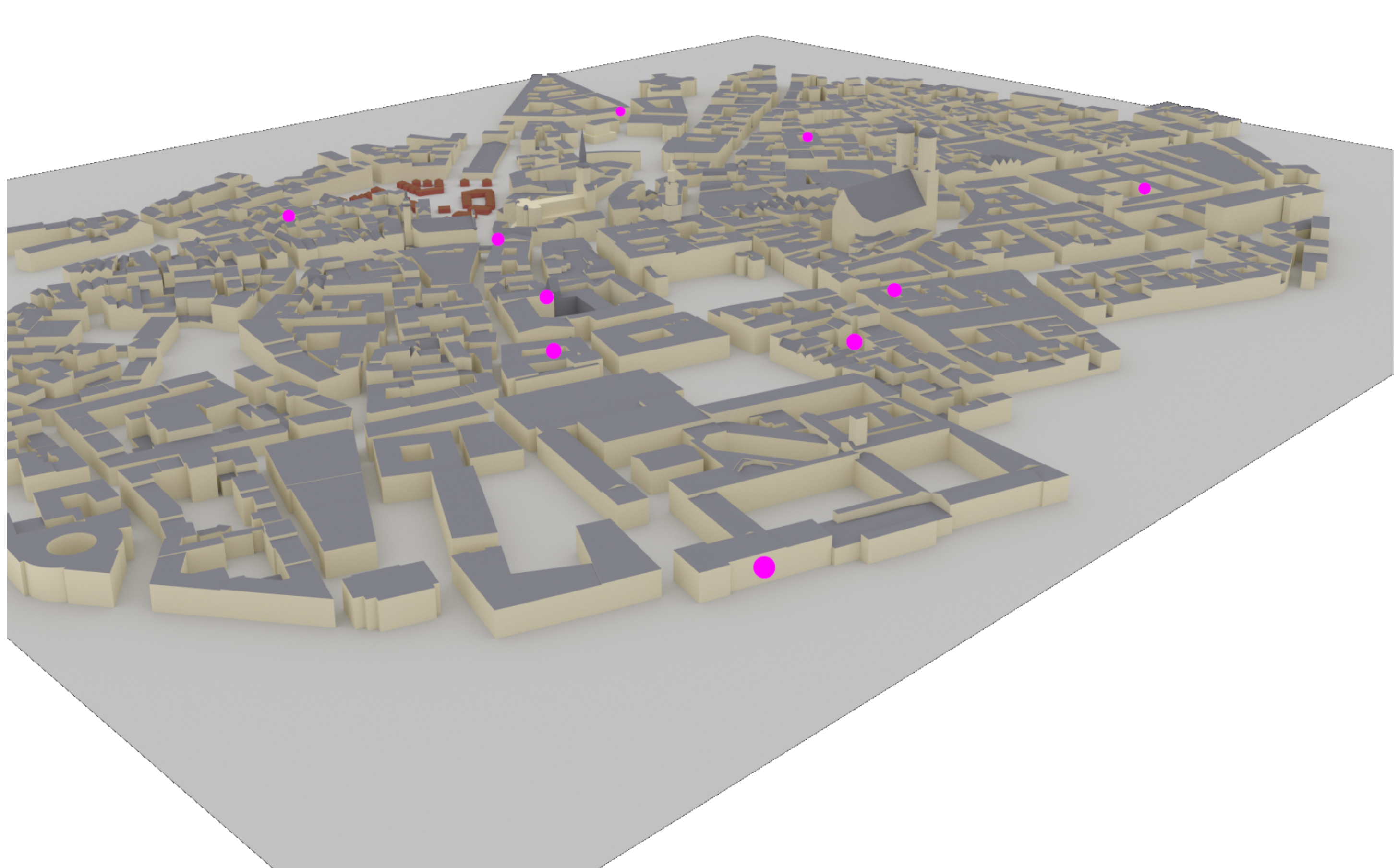}
        \label{fig:tx_locations_munich}
    \end{subfigure}
    \hspace{0.01\textwidth}
    \begin{subfigure}{0.44\textwidth}
        \centering
        \includegraphics[width=\linewidth]{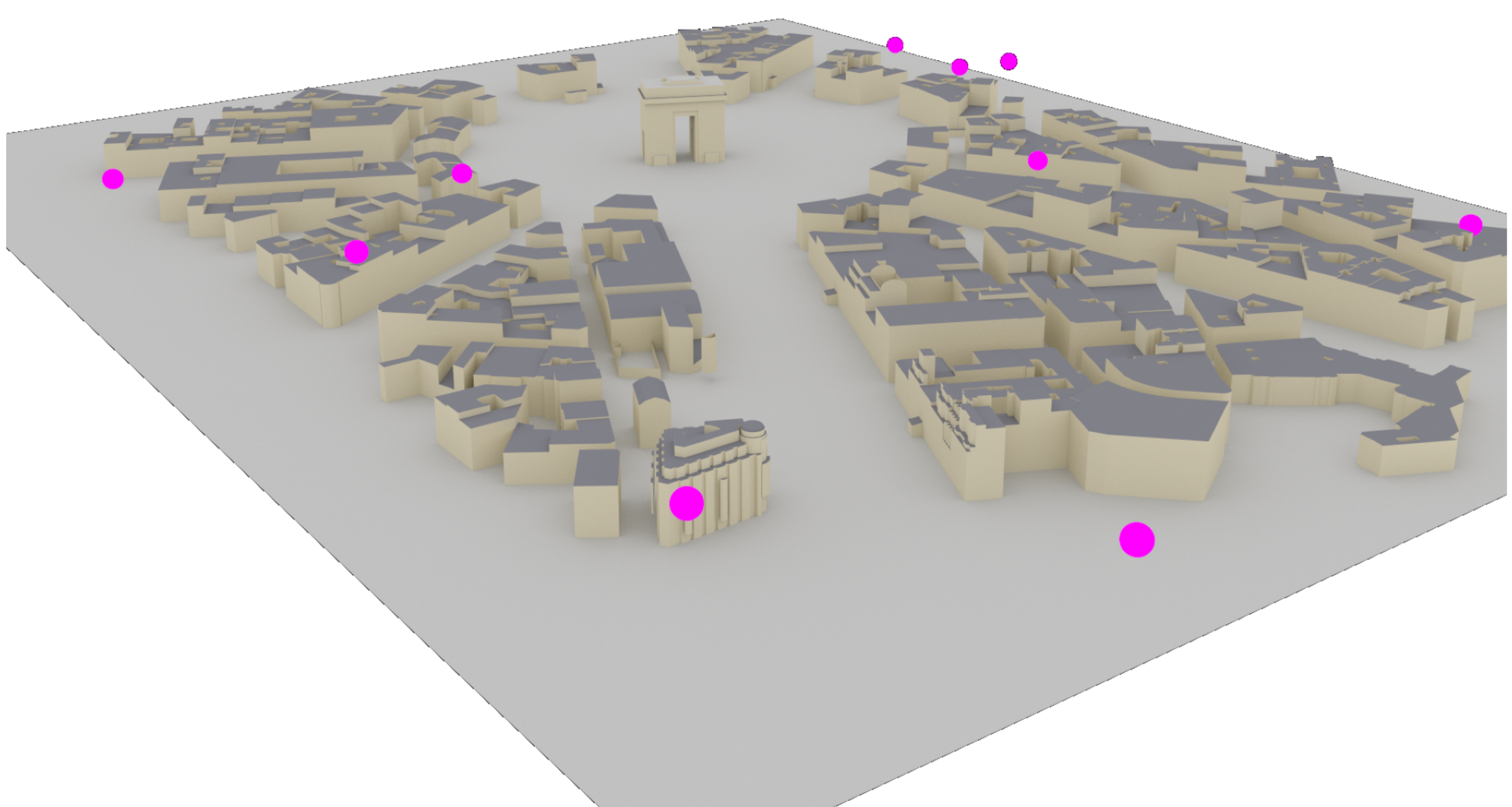}
        \label{fig:tx_locations_etoile}
    \end{subfigure}
    \caption{Transmitter Locations Used for Sensitivity Analysis on the Munich and Etoile Scenes}
    \label{fig:combined_broken_links}
\end{figure*}


\subsection{Background}
The precision of digital spectrum twins heavily depends on the accuracy of environmental geometry data, including building structures and terrain features. Several studies have highlighted the significant impact that inaccuracies in urban geometry databases can have on propagation predictions.

Early research by \cite{accuracy_of_env_representation} emphasized that errors in building shapes and positions lead to notable prediction inaccuracies in propagation models. However, these inaccuracies were not fully quantified or systematically analyzed.

To explore the influence of map inaccuracies, \cite{influence_of_db_acc} based their propagation predictions on outdated two-dimensional (2D) building layouts derived from various map types. They introduced artificial random errors by displacing building walls and corners using normal distributions. While this approach highlighted the limitations of using obsolete maps and unconventional perturbation techniques, it did not accurately reflect realistic urban changes. 

Further investigating the sensitivity of ray-tracing models, \cite{investigation_rt} examined factors such as wall characteristics, antenna position offsets, and inaccuracies in building databases. Their findings indicated that displacements greater than one meter can significantly degrade model performance. Notably, receiver positions at the boundaries of line-of-sight (LOS) areas and in deep shadow regions experienced the highest power deviations.

Studies focusing on indoor environments have also provided valuable insights into the effects of geometric inaccuracies. \cite{geo_perturb} modeled lateral wall positions as random variables to investigate their impact on ray-tracing results. They found that while received power is significantly affected by wall position uncertainties, the rms delay spread remains largely insensitive. However, their analysis was limited to controlled indoor settings and did not extend to the complexities of urban outdoor environments.

In efforts to enhance prediction accuracy through material properties, \cite{Jemai2009} improved deterministic channel models by calibrating optimal material parameters using pilot measurements and simulated annealing techniques. Although this method enhanced accuracy across various positions, it relied on manual calibration and extensive measurements. 

Another indoor study by \cite{indooreffects} examined the sensitivity of ray-tracing results to inaccuracies in the dielectric constants of building materials. By treating these constants as random variables and employing Monte Carlo analysis, they observed that path gain remains relatively unaffected, whereas rms delay spread is significantly sensitive to material property uncertainties. This emphasizes the importance of accurate electromagnetic parameters but also highlights the limitations of focusing solely on indoor environments.

Extending the focus to outdoor settings, \cite{opt_for_perm_cond} investigated the impact of adjusting permittivity values for building walls, roofs, and street floors in 3D ray-tracing models. Their study revealed that fine-tuning these material properties improves the accuracy of path loss estimations despite the heterogeneity of real-world materials.

Efforts to improve geometry extraction methods have also been explored. \cite{geospatial_resources} proposed extracting 3D building structures from single-view internet images by combining height measurements with 2D projective transformations. While this method demonstrated accurate path-loss calculations in ray-tracing models, it also revealed that inaccuracies in building footprints and heights can impact path-loss predictions. Despite errors remaining within typical ranges between experimental measurements and simulations, relying on single-view images may not capture the full complexity of urban environments, highlighting the need for more robust geometry extraction techniques.

Collectively, these studies highlight the importance of accurate environmental geometry data and material properties in propagation modeling. However, limitations persist, including:

\begin{itemize}
\item 
Reliance on outdated or incomplete data sources: Using obsolete maps or limited views may not accurately represent current urban environments.

\item 
Focus on controlled indoor environments: Many studies do not address the variabilities inherent in urban outdoor scenarios.

\item
Limited sensitivity analyses: There is a need for more comprehensive investigations into how uncertainties in geometry and material properties affect key channel statistics both individually and collectively.

\end{itemize}

Our work addresses these gaps by introducing an automated geometry extraction process designed for ray-tracing propagation models, specifically utilizing the NVIDIA Sionna Ray-Tracing tool \cite{sionnart}. By leveraging multiple data sources including OpenStreetMap \cite{openstreetmap}, the Microsoft Global ML Building Footprints dataset \cite{globalmlbuildingfootprints} for building data, and the U.S. Geological Survey (USGS) \cite{usgs} for terrain data we enhance the accuracy and scalability of environmental representations.

Additionally, we present an open-source implementation of a building geometry and material perturbation pipeline. This pipeline introduces Gaussian noise to building heights, positions, relative permittivity, and conductivity, enabling a systematic exploration of how uncertainties in these parameters affect propagation modeling. Our extensive sensitivity analysis examines the impact of these perturbations on key channel statistics such as path gain, mean excess delay, delay spread, and link reliability. We also investigate the co-variation of these statistics under identical perturbations and their variation with different transmitter-receiver separation distances.

\section{Methodology}
\subsection{Geometry Extraction}
The geometry extraction process for ray-tracing propagation models is automated through a pipeline depicted in Figure \ref{fig:ge_pipeline} that transforms user-specified region corner coordinates into detailed 3D scenes suitable for electromagnetic simulations. This automation enhances efficiency and consistency in creating models by reducing manual intervention. There are two primary workflows in this pipeline, each integrating different data sources and tools to generate accurate geometries.

In the first workflow, the process begins with the user specifying the corner coordinates of the desired outdoor region. The Blender Python API is employed to automate tasks within Blender, an open-source 3D modeling software. Using this API, the Blosm add-on is activated to interface with OpenStreetMap (OSM), from which it retrieves comprehensive building footprint data for the specified region. This building data is imported directly into Blender as meshes. Subsequently, terrain elevation data is obtained by Blosm from a GitHub repository that aggregates information from various sources covering numerous countries. In the U.S., this data is provided by the 3D Elevation Program of the United States Geological Survey (USGS). The terrain data is then imported into Blender and converted into a mesh to represent the region's topography, achieving a resolution of 10 meters in the U.S. (excluding Alaska) and varying resolutions in other countries.

Within Blender, the building and terrain meshes are refined to ensure they meet the requirements for ray-tracing simulations. This includes assigning ITU materials to objects and arbitrarily selecting colors for different materials. After the meshes are prepared, the Mitsuba-Blender add-on is used to export the entire scene as an XML file. The final step involves importing the XML file and associated meshes into NVIDIA Sionna RT \cite{sionnart}, a framework for simulating RF propagation using ray-tracing techniques. Three example scenes generated using this automated geometry extraction pipeline are displayed in Figure \ref{fig:scenes_combined}.

The second workflow offers an alternative approach by bypassing the Blosm add-on and instead using external data sources and Python scripting for mesh creation. One potential advantage of this alternative workflow is its ability to provide a more accurate representation of rural buildings without relying on OpenStreetMap, which is often infrequently updated in rural areas. As in the first workflow, the user provides the corner coordinates of the target region to define the area of interest. Building data is sourced from the Microsoft Global ML Building Footprints repository \cite{globalmlbuildingfootprints}, which offers detailed building structures derived from satellite imagery processed through computer vision algorithms. High-resolution terrain data is again obtained from the USGS digital elevation maps this time with 1-m resolution.

Custom Python scripts are developed to process the building and terrain data, generating triangular meshes without the need for Blender's Blosm add-on. Once the meshes are generated, they are imported into Blender for consolidation and any additional adjustments including material and color assignments to objects. Blender serves as a platform to unify the building and terrain meshes and to leverage the Mitsuba-Blender add-on for scene export. The generated XML file along with the meshes can now be imported into Sionna RT for ray-tracing simulations.

\subsection{Sensitivity Analysis}
In the second part of the work, a comprehensive ray-tracing sensitivity analysis was conducted using two representative urban scenes provided in Sionna RT: the Munich and Etoile scenes. Within each scene, ten transmitter (Tx) locations were selected at random. For each Tx location, the original scene geometry and material properties were systematically perturbed 50 times, producing a series of modified environments. In these perturbed environments, numerous receiver (Rx) positions were evaluated, and the resulting channel characteristics were extracted and statistically analyzed. The channel statistics considered included path gain, mean excess delay, delay spread, link outage, and the Rician \textit{K}-factor. For each of these statistics, the standard deviation across the 50 perturbations was computed to assess sensitivity.

Perturbations were introduced to reflect plausible uncertainties in geometric and material parameters. For geometric perturbations in building heights, Gaussian noise with a standard deviation ($\sigma$) of 1 meter was added to the z-coordinates of the buildings’ upper vertices. Similarly, building positions were perturbed by adding Gaussian noise with $\sigma = 0.4$ meters to the x and y coordinates of the vertices. Material properties, such as relative permittivity and conductivity, were also altered to simulate uncertainty in electromagnetic parameters. Assuming that initial material parameters were approximately correct, Gaussian noise with a standard deviation equal to 10\% of the initial values was added.

The scene and perturbation configuration parameters are summarized in Table \ref{tab:scene_config}.

\begin{table}[ht]
\centering
\setlength{\tabcolsep}{8pt} 
\begin{tabular}{p{3.8cm} p{4cm}}
\toprule
\textbf{Parameters} & \textbf{Values} \\
\midrule
Number of Unique Tx Locations \\ per Scene & 10\\ 
\specialrule{0.4pt}{0pt}{0pt}
Number of Perturbations per \\ Type of Perturbation & 50\\ \specialrule{0.4pt}{0pt}{0pt}
Considered Scenes & Munich, Etoile\\ \specialrule{0.4pt}{0pt}{0pt}
Types of Perturbations & Material, Height, Position, Height \& Position, Material, Height \& Position \\  \specialrule{0.4pt}{0pt}{0pt} 
Frequency of Operation & 3.5 GHz\\ \specialrule{0.4pt}{0pt}{0pt}
Tx \& Rx Radiation Patterns & Isotropic \\ \specialrule{0.4pt}{0pt}{0pt} 
Tx \& Rx Antenna Polarization & Vertical\\ 
 \specialrule{0.4pt}{0pt}{0pt} 
Number of Tx \& Rx Elements & 1\\
 \specialrule{0.4pt}{0pt}{0pt} 
Std. Dev. of Gaussian Noise Added to Building Heights & 1 m\\
 \specialrule{0.4pt}{0pt}{0pt} 
Std. Dev. of Gaussian Noise Added to Building Positions & 0.4 m\\
 \specialrule{0.4pt}{0pt}{0pt} 
Std. Dev. of Relative Permittivity Added to Building Materials & 10 \% of the Initial Value\\
 \specialrule{0.4pt}{0pt}{0pt} 
Std. Dev. of Conductivity Added to Building Materials & 10 \% of the Initial Value\\
 \specialrule{0.4pt}{0pt}{0pt} 
Coverage Map \& Path Configuration & Line-of-sight (LOS), up to 5 reflections, diffraction and edge diffraction, no scattering, 1 million samples\\

\bottomrule
\end{tabular}
\caption{Scene \& Perturbation Configuration Parameters}
\label{tab:scene_config}
\end{table}

\begin{figure*}[htbp]
    \centering
    \begin{subfigure}[t]{0.43\textwidth}
        \centering
        \includegraphics[width=\linewidth]{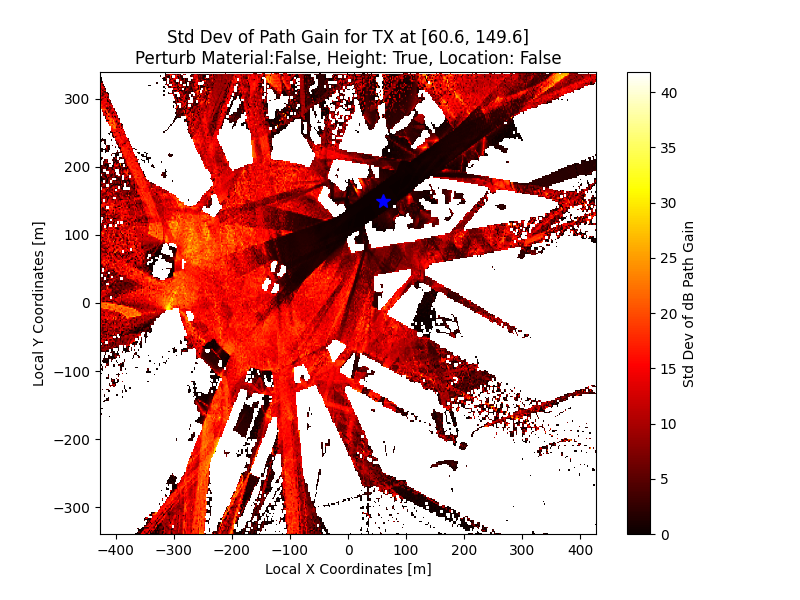}
        \caption{Path Gain Standard Deviation, Height Perturbation, Etoile}
        \label{fig:pg_std_tx_etoile}
    \end{subfigure}
    \hspace{0.005\textwidth} 
    \begin{subfigure}[t]{0.43\textwidth} 
        \centering
        \includegraphics[width=\linewidth]{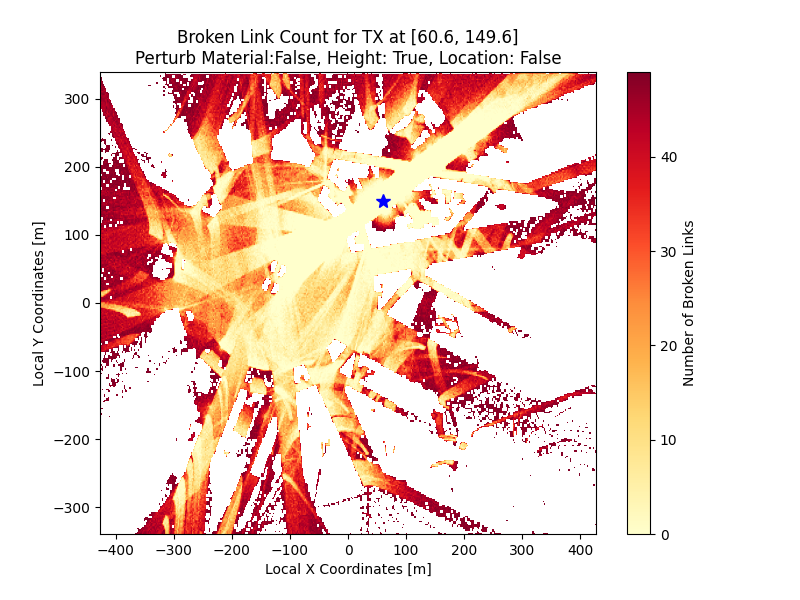}
        \caption{Link Outage Frequency, Height Perturbation, Etoile}
        \label{fig:link_outage_freq_tx_etoile}
    \end{subfigure}
    
    \vspace{0.1cm}
    
    \begin{subfigure}[t]{0.43\textwidth}
        \centering
        \includegraphics[width=\linewidth]{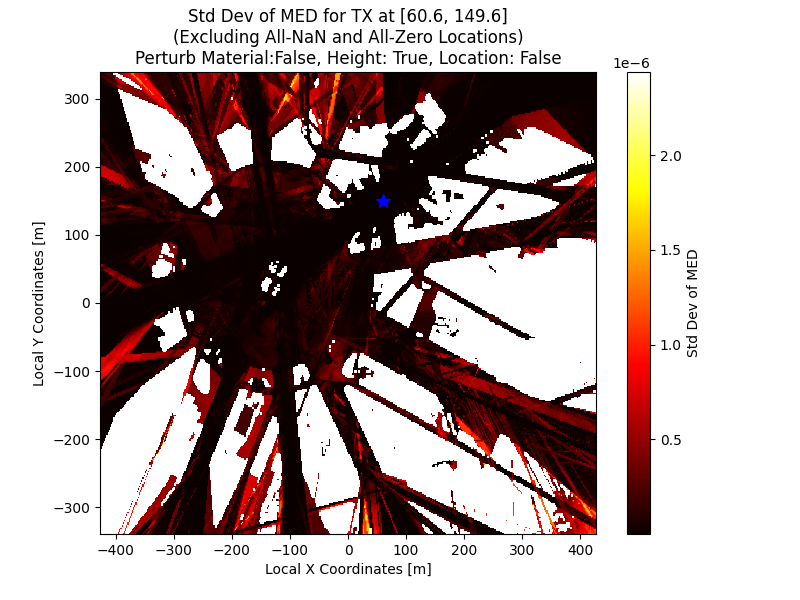}
        \caption{Mean Excess Delay Standard Deviation, Height Perturbation, Etoile}
        \label{fig:med_std_tx_etoile}
    \end{subfigure}
    \hspace{0.005\textwidth} 
    \begin{subfigure}[t]{0.43\textwidth}
        \centering
        \includegraphics[width=\linewidth]{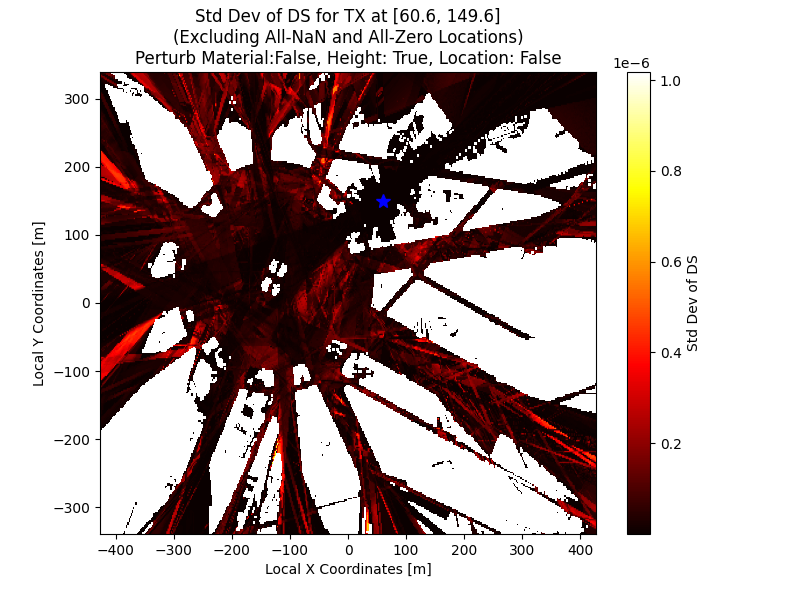}
        \caption{Delay Spread Standard Deviation, Height Perturbation, Etoile}
        \label{fig:ds_std_tx_etoile}
    \end{subfigure}
    
    \caption{Analysis of Path Gain, Mean Excess Delay, and Delay Spread Standard Deviations and Link Outage Frequency with Height Perturbation in Etoile Scene}
    \label{fig:etoile_height_analysis}
\end{figure*}

\section{Results \& Analysis}

This section presents a detailed examination of how environmental perturbations, specifically building height, position, and material properties, impact channel statistics in complex urban environments. We first focus on how channel metrics vary under height perturbations, then analyze the interrelationships among the metrics under all types of perturbations combined. Finally, we compare the effects of different perturbation types, highlight scene-specific observations, and discuss the implications of these results for accurate propagation modeling and DST development.

\subsection{Influence of Height Perturbations on Key Channel Metrics}

\subsubsection{Path Gain Variability and Distance-Dependent Effects}

Figure \ref{fig:etoile_height_analysis} reveals important insights into how height perturbations affect certain channel statistics, including path gain (PG), link outage frequency, mean excess delay (MED), and delay spread (DS). Specifically, Figure \ref{fig:pg_std_tx_etoile} shows that the standard deviation of path gain becomes especially large when the link between the transmitter and receiver is not achievable through line-of-sight (LOS) or multiple layers of reflections. In such cases, diffraction and edge diffraction are the primary contributors to path gain and are significantly affected by changes in building heights which in turn makes the path gain standard deviation large at those locations. The overall effect is visible through dark traces of reflection rays and a dark corridor that is in direct LOS with the transmitter, contrasted by brighter diffraction regions.  However, when the distance from the transmitter exceeds a certain threshold, even receivers with a direct LOS begin to experience significant deviations. This happens because the power of the LOS ray decreases with distance, making the overall received power more dependent on reflection and diffraction rays. Although reflections near the transmitter are less affected by height perturbations, at larger distances, reflection rays are influenced considerably. This is due to the cumulative effect of height changes on multi-layer rays that undergo numerous reflections and diffractions. As the distance from the transmitter increases even further, the variance reduces potentially due to "left-censored data" \cite{leftcensoring} because powers lower than the receiver sensitivity threshold are not included in variance calculation and the link is considered to be broken as also encountered in \cite{georeferenced_spec_occ}.

\begin{figure*}[htbp]
    \centering
    \begin{subfigure}{0.49\textwidth}
        \centering
        \includegraphics[width=\linewidth]{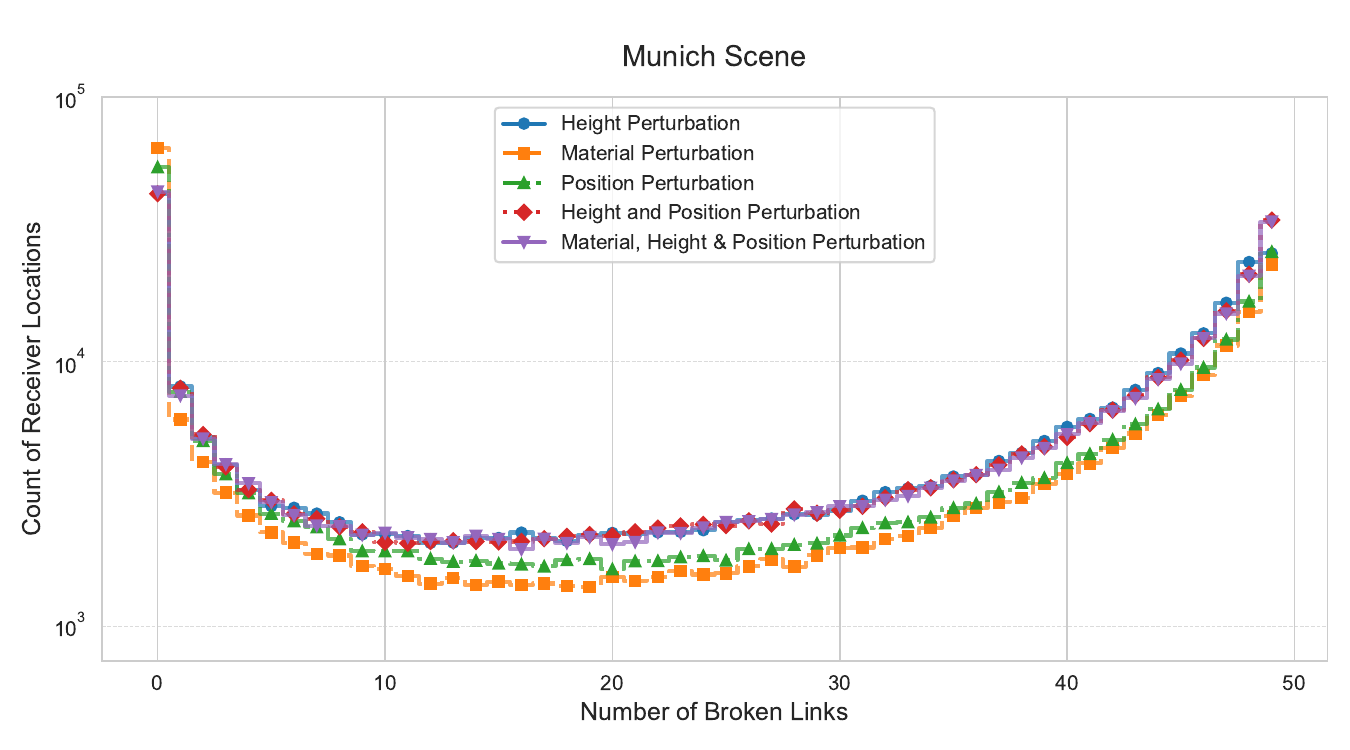}
        \caption{Munich Perturbations}
        \label{fig:broken_links_munich}
    \end{subfigure}
    \hfill
    \begin{subfigure}{0.49\textwidth}
        \centering
        \includegraphics[width=\linewidth]{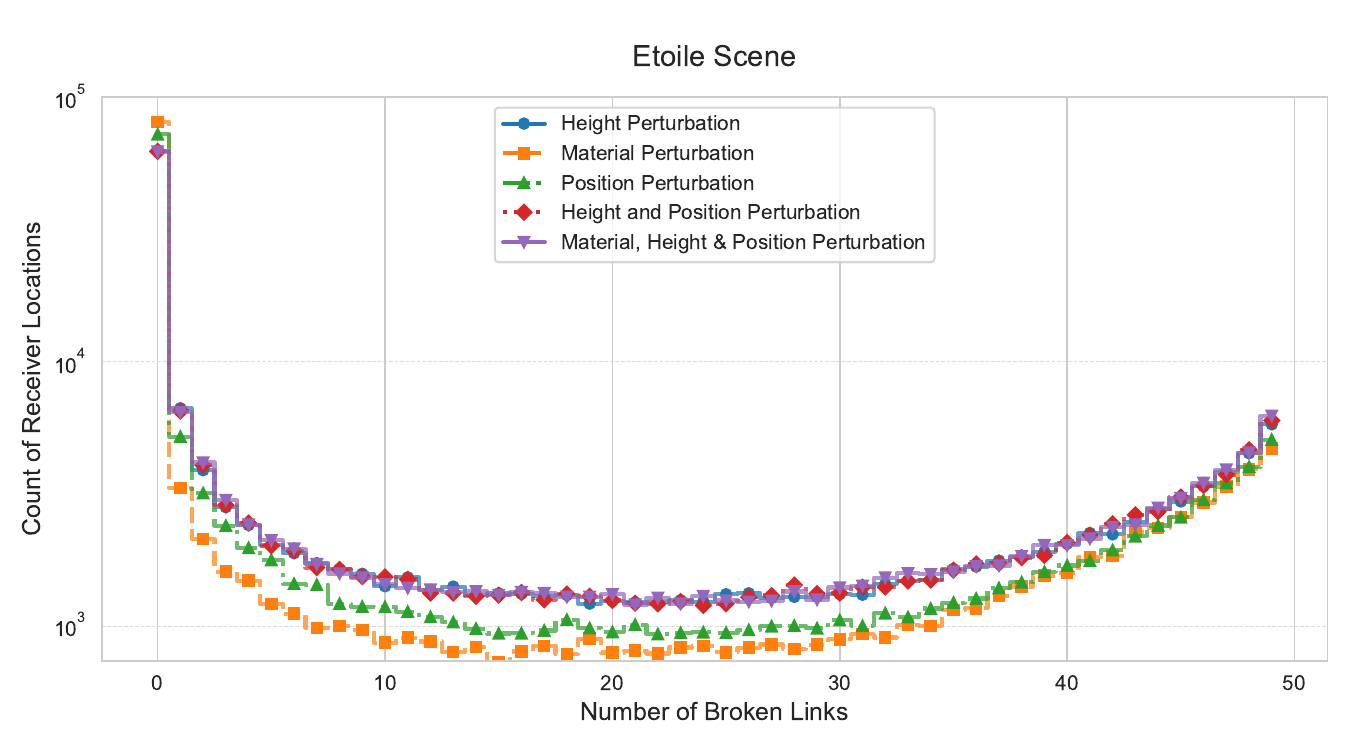}
        \caption{Etoile Perturbations}
        \label{fig:broken_links_etoile}
    \end{subfigure}
    \caption{Histograms of Broken Links for Different Types of Perturbations in Munich and Etoile.}
    \label{fig:combined_broken_links}
\end{figure*}


\subsubsection{Link Outage Frequency and Height Sensitivity}

Some areas with frequent link outages, as shown in Figure \ref{fig:link_outage_freq_tx_etoile}, correlate with regions containing shorter buildings. Under significant (though unlikely) negative height perturbations, these buildings may no longer block rays, occasionally allowing a weak connection to form beneath the fixed receiver height of 1.5 meters. This rare reconnection inflates the observed outage frequency. 

The resulting U-shaped distribution of link outages, seen in Figures \ref{fig:broken_links_munich} and \ref{fig:broken_links_etoile}, reflects these threshold effects. Large outage segments occur where substantial height changes open or close paths that are normally absent due to building obstruction or distance-related signal weakening. In these zones, even small vertical adjustments to building heights critically influence whether any signal reaches the receiver.

\subsubsection{Mean Excess Delay (MED) and Height-Driven Threshold Effects}

If a building is just tall enough in its original configuration to block a near LOS path, lowering its height may allow that path to slip through, shortening the MED. Conversely, increasing building height might introduce additional blocking of previously available multipath routes, changing the structure of the delay profile.
Such threshold-like effects are particularly pronounced when receivers are near boundaries that determine path availability. These boundary conditions make MED highly sensitive to height perturbations.

Locations that depend primarily on ground-level reflections or are in unobstructed LOS to the transmitter are much less sensitive to vertical changes in building structure. In those areas, modifying building heights slightly does not significantly alter the primary arrival time or the set of dominant multipath components. This is visible in Figure \ref{fig:med_std_tx_etoile}.

On the other hand, environments dominated by complex vertical scattering mechanisms such as reflections from slanted rooftops, diffraction around building crowns, or energy guided along elevated canyons are inherently more "vertically sensitive". Such regions amplify the effect of small height perturbations, resulting in higher MED variability as observed at bright locations in Figure \ref{fig:med_std_tx_etoile}.

\begin{figure*}[htbp]
    \centering
    \begin{subfigure}[t]{0.32\textwidth}
        \centering
        \includegraphics[width=\linewidth]{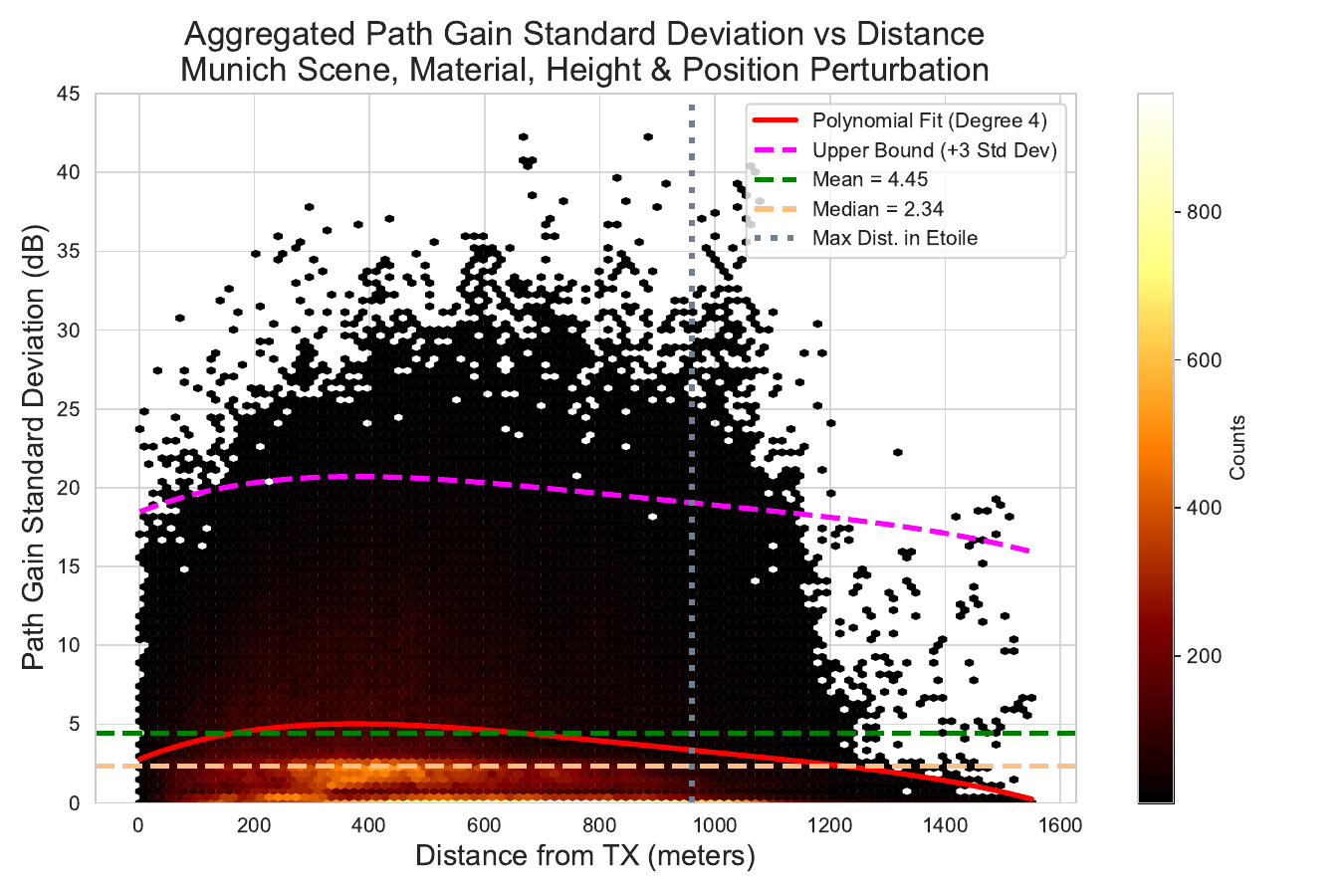}
        \caption{Path Gain Standard Deviation vs. Distance, Munich}
        \label{fig:pg_std_munich}
    \end{subfigure}
    \hfill
    \begin{subfigure}[t]{0.32\textwidth}
        \centering
        \includegraphics[width=\linewidth]{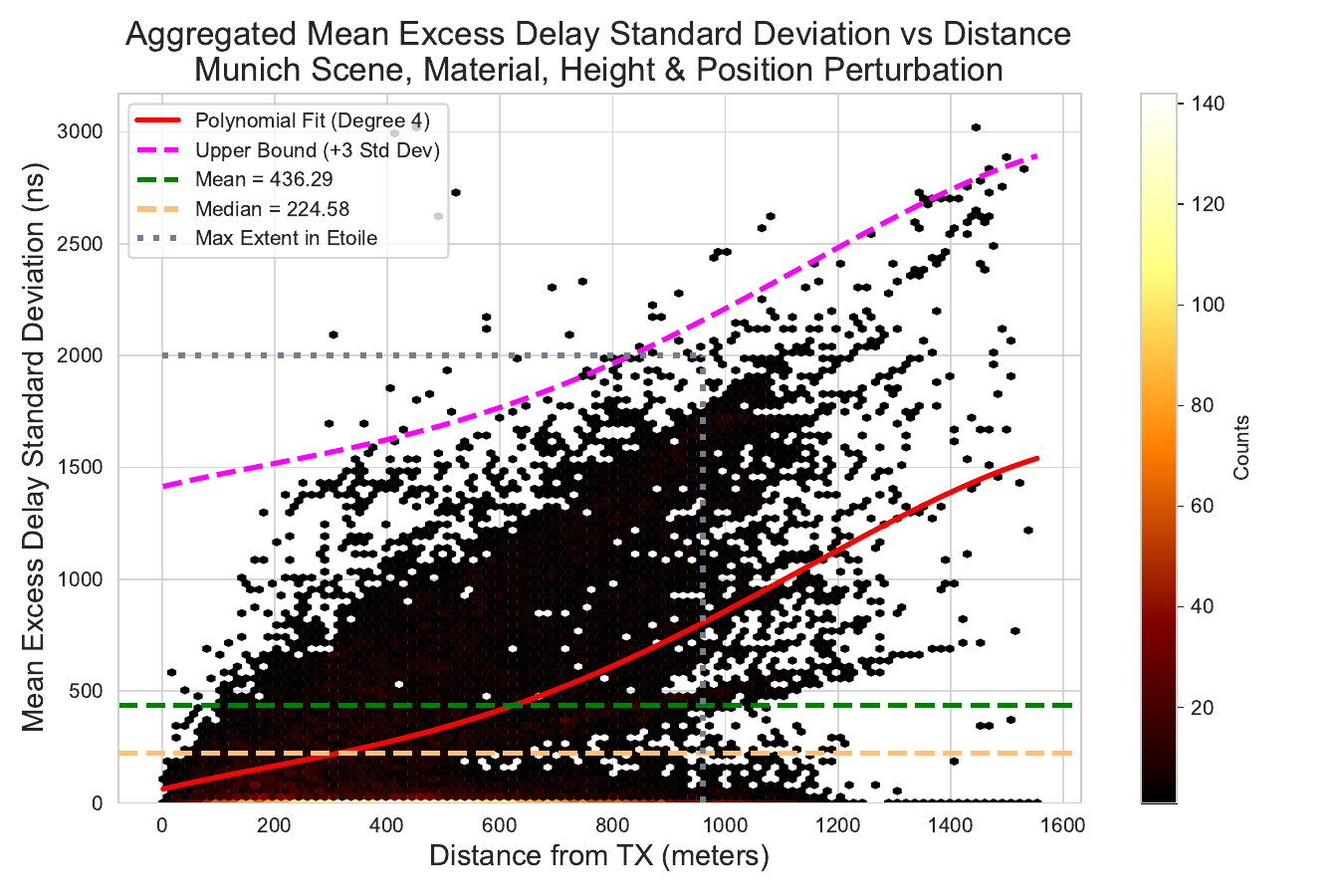}
        \caption{Mean Excess Delay Standard Deviation vs. Distance, Munich}
        \label{fig:med_std_munich}
    \end{subfigure}
    \hfill
    \begin{subfigure}[t]{0.32\textwidth}
        \centering
        \includegraphics[width=\linewidth]{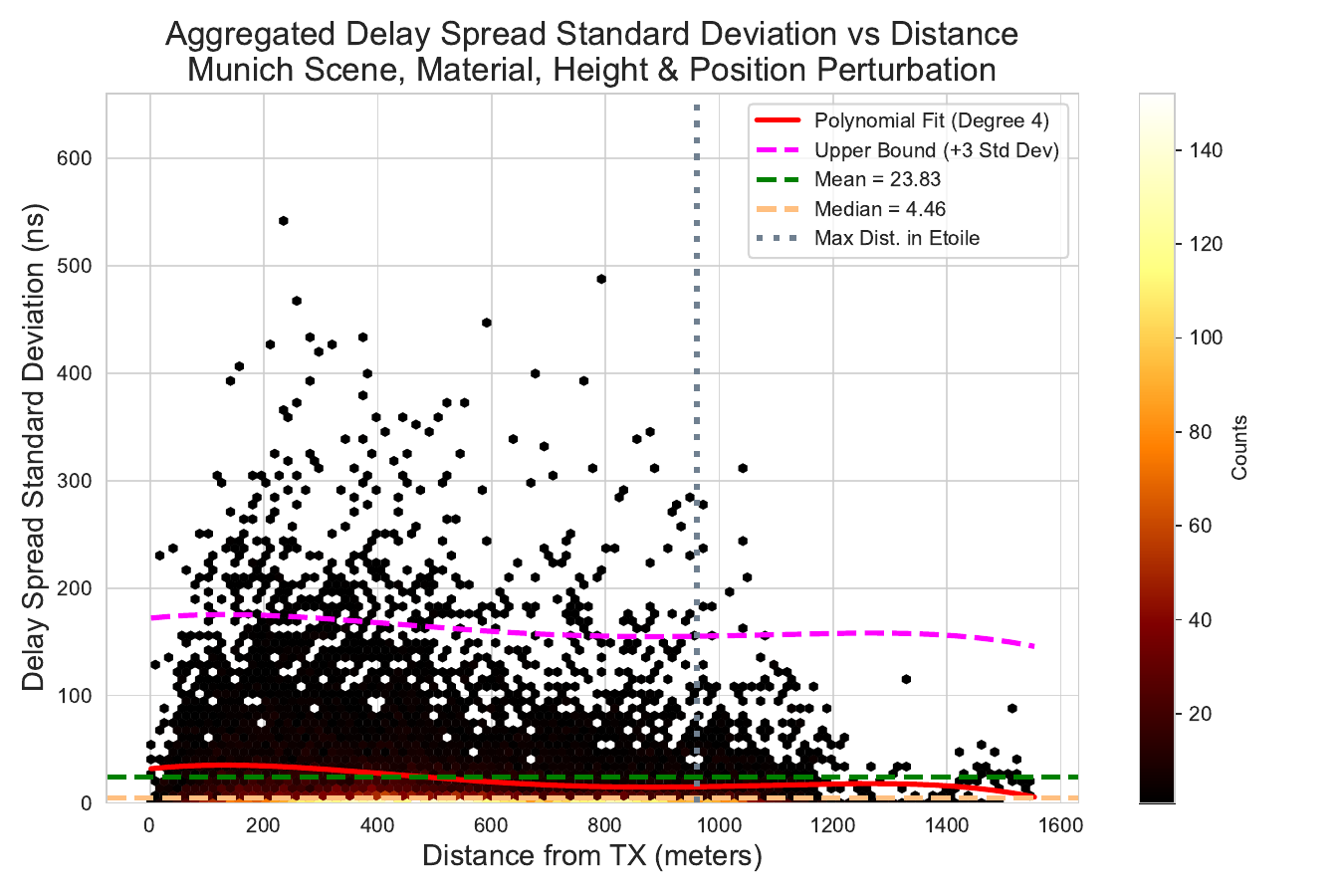}
        \caption{Delay Spread Standard Deviation vs. Distance, Munich}
        \label{fig:ds_std_munich}
    \end{subfigure}
    
    \vspace{0.4cm}
    
    \begin{subfigure}[t]{0.32\textwidth}
        \centering
        \includegraphics[width=\linewidth]{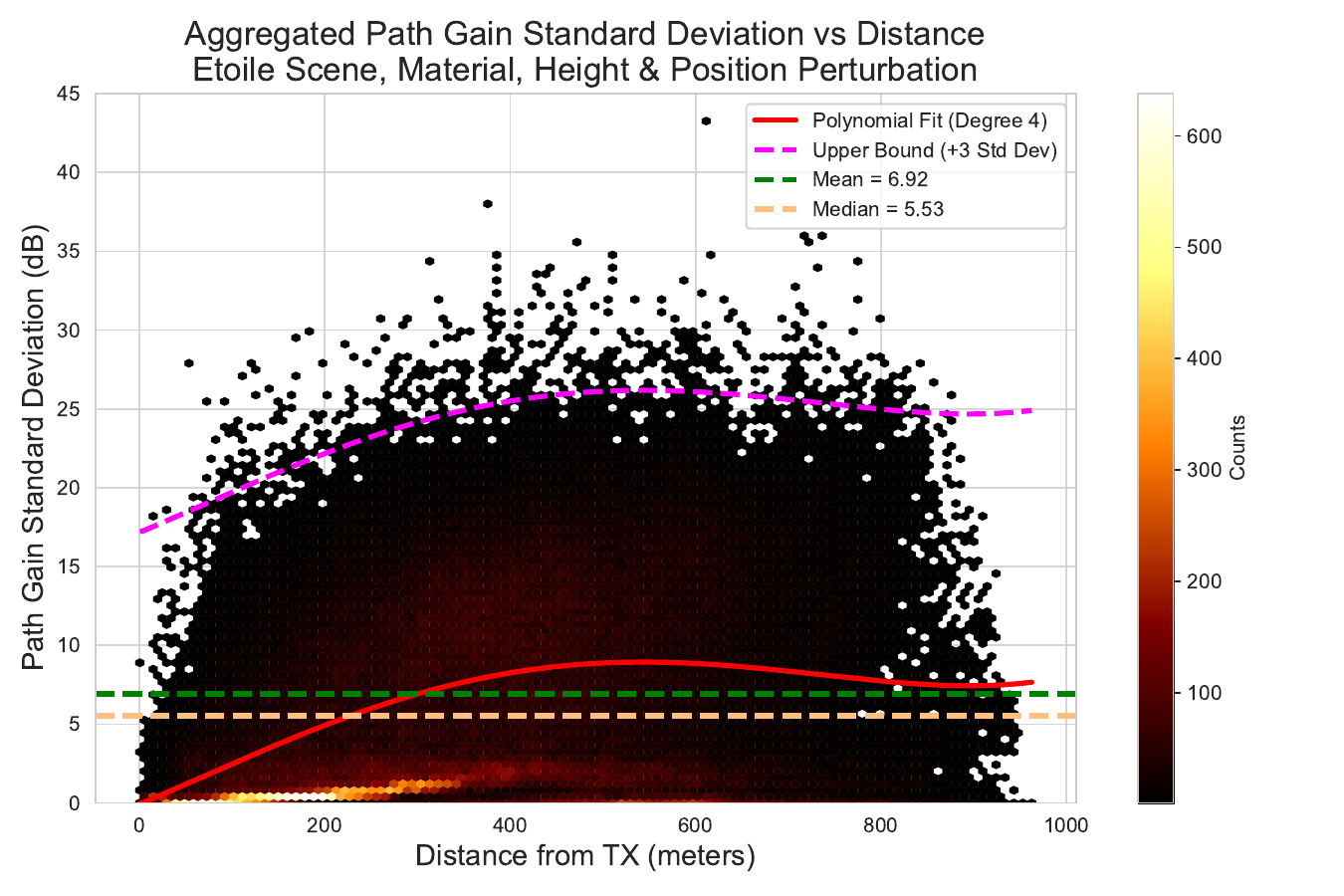}
        \caption{Path Gain Standard Deviation vs. Distance, Etoile}
        \label{fig:pg_std_etoile}
    \end{subfigure}
    \hfill
    \begin{subfigure}[t]{0.32\textwidth}
        \centering
        \includegraphics[width=\linewidth]{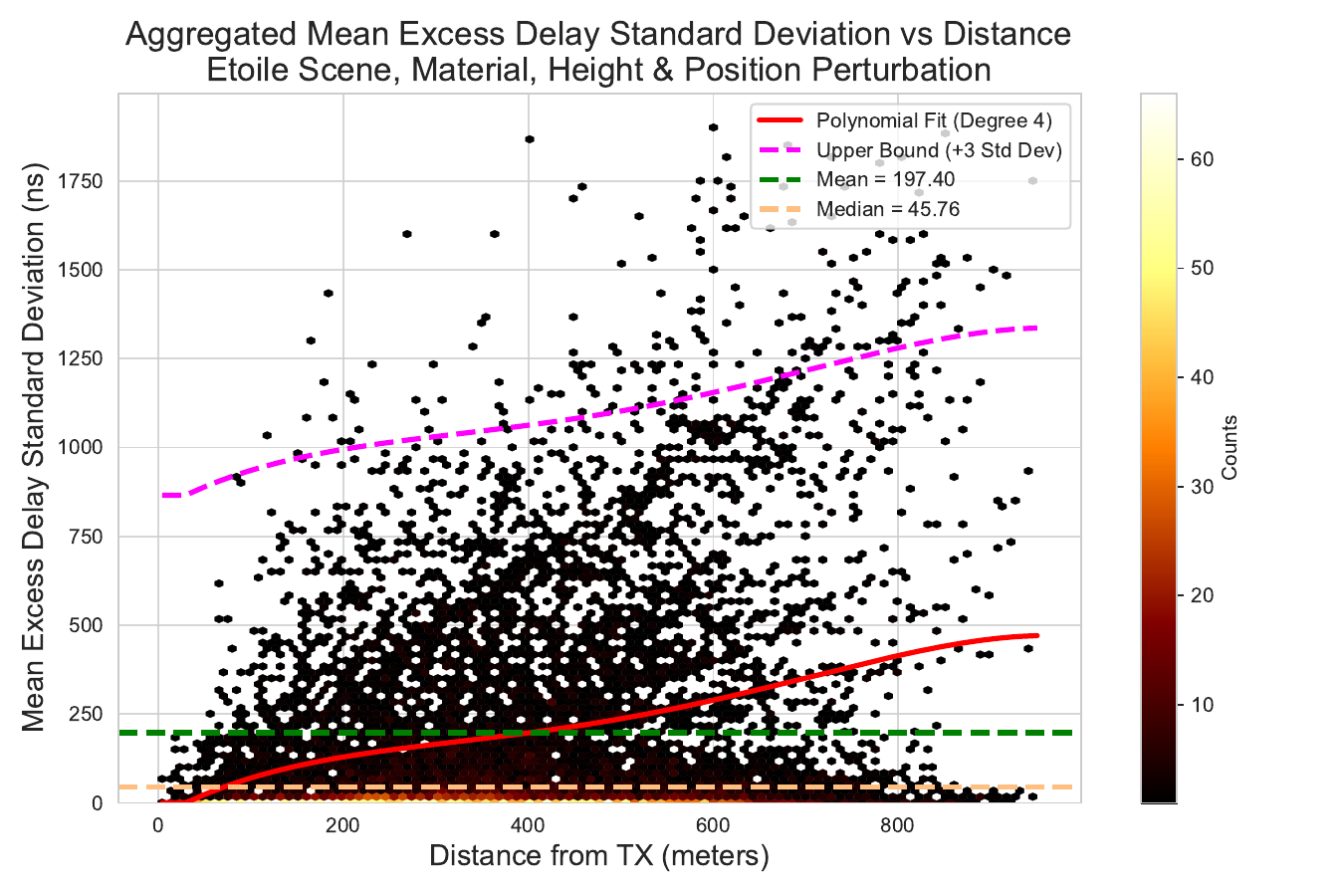}
        \caption{Mean Excess Delay Standard Deviation vs. Distance, Etoile}
        \label{fig:med_std_etoile}
    \end{subfigure}
    \hfill
    \begin{subfigure}[t]{0.32\textwidth}
        \centering
        \includegraphics[width=\linewidth]{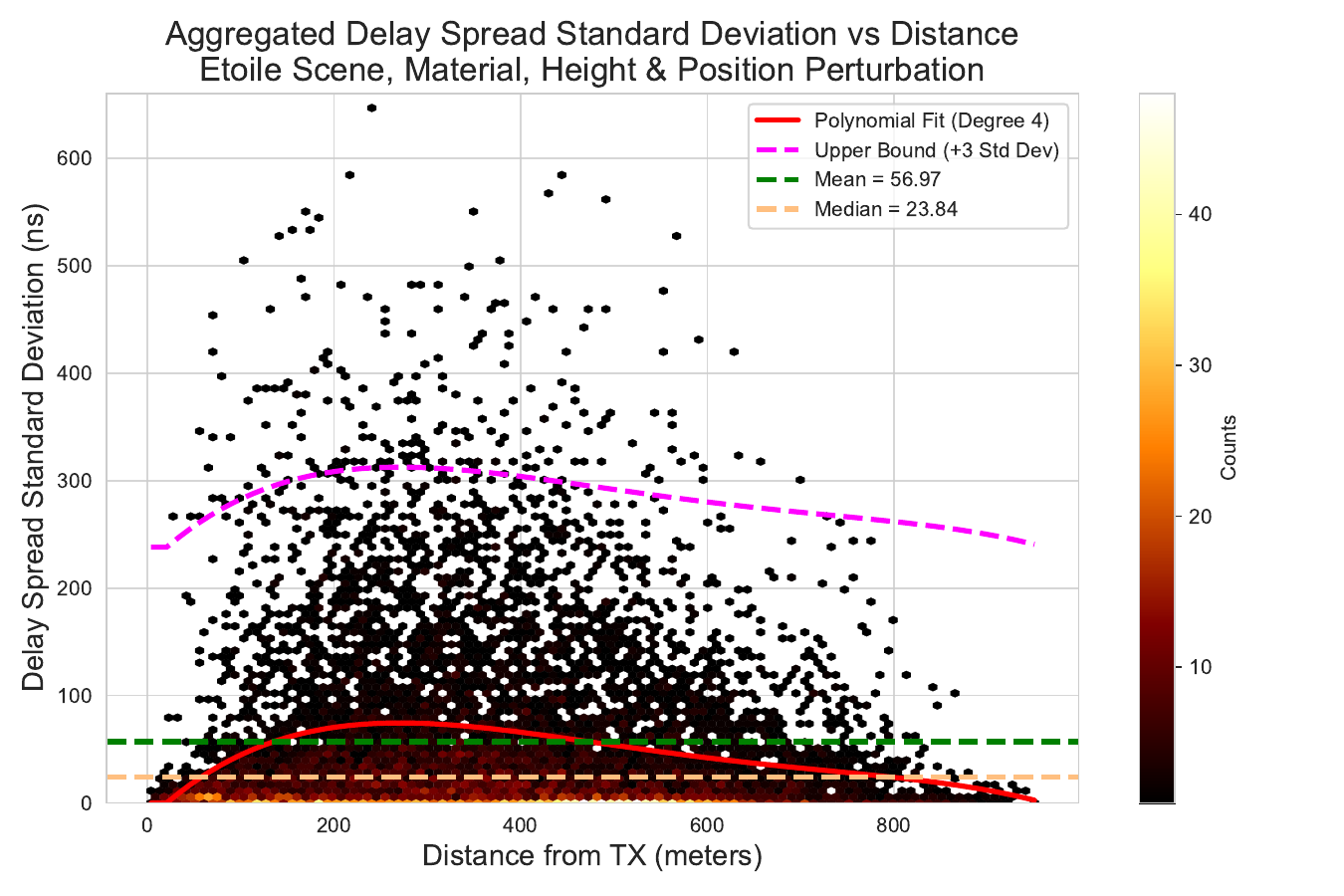}
        \caption{Delay Spread Standard Deviation vs. Distance, Etoile}
        \label{fig:ds_std_etoile}
    \end{subfigure}
    
    \caption{Combined Analysis of Path Gain, Mean Excess Delay, and Delay Spread Standard Deviations for Munich and Etoile Across All Perturbation Types}
    \label{fig:combined_analysis}
\end{figure*}

The clear, positive MED variability trend with increasing distance as depicted in Figures \ref{fig:med_std_munich} and \ref{fig:med_std_etoile} is typically due to the signal traversing more diverse and heterogeneous urban terrain at longer distances. Different building materials, varying densities of structures, and changes in topography all factor into the complexity of multipath generation. When uncertainties such as slight variations in building height are introduced, their effects can be magnified as the signal has passed through a richer set of scatterers and obstacles. This accumulated complexity increases the sensitivity of delay characteristics.

\subsubsection{Delay Spread (DS) and Limited Complexity Growth}

The DS standard deviations depicted in Figures \ref{fig:ds_std_munich} and \ref{fig:ds_std_etoile} however do not follow the same monotonically increasing trend with distance. Without scattering, and with a cap of five reflections per ray, the complexity of multipath propagation is inherently limited in the simulations. At very short distances, the absence of scattering and few low-order reflections yield a stable delay spread, resulting in minimal sensitivity to building height perturbations. As distance increases, the finite number of allowed reflections temporarily boosts complexity, making the delay spread more sensitive and possibly increasing its variability. However, because there are no scattered paths and a strict reflection limit, this complexity cannot grow indefinitely. At larger distances, many potential multipath components are either too weak or geometrically unavailable, causing the DS standard deviation to plateau or taper off rather than continue increasing.

\subsection{Interrelationships Among Channel Metrics}

\begin{figure*}
    \centering
    \includegraphics[width=\linewidth]{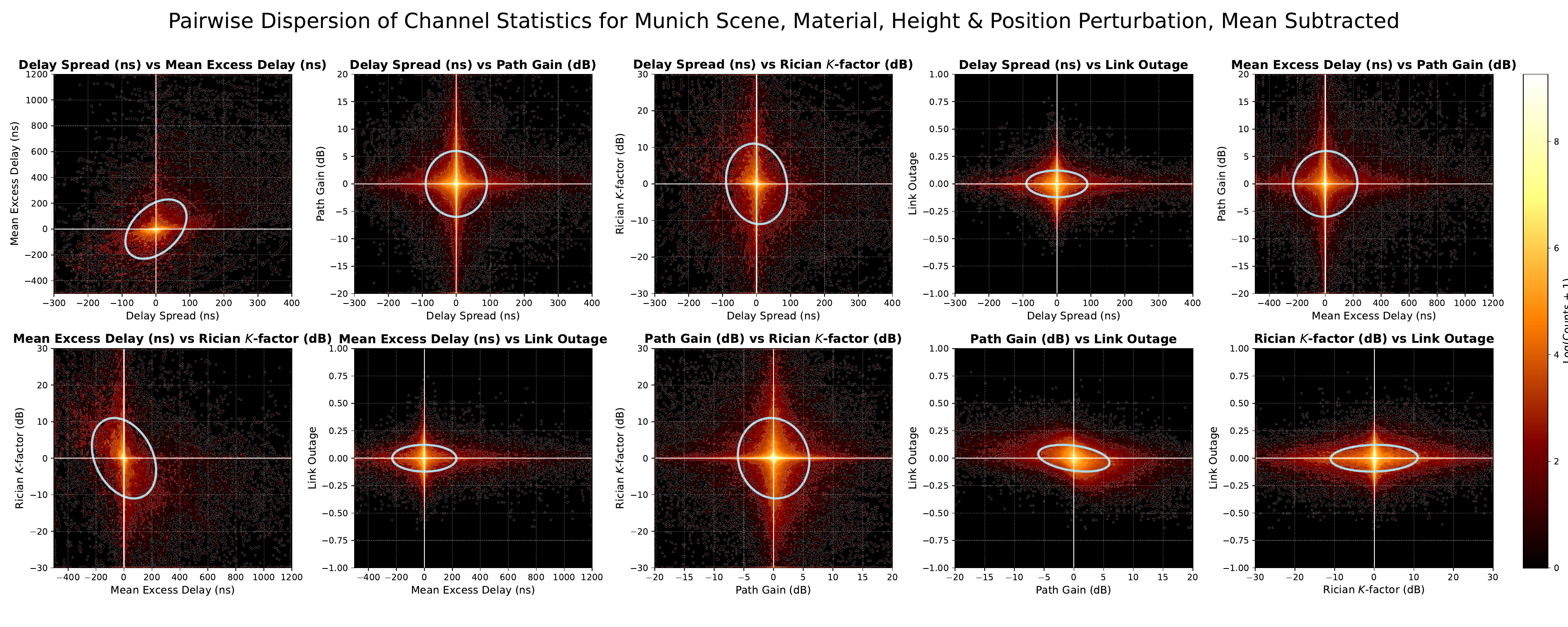}
    \caption{Pairwise Dispersion of Channel Metrics for Combined Perturbations of All Types in the Munich Scene}
    \label{fig:dispersion_munich}
\end{figure*}

Another important analysis is to examine how the channel statistics considered vary in response to perturbations. Figure \ref{fig:dispersion_munich} displays the aggregate dispersion plots across all transmitter locations for all pairs of channel statistics. These include path gain, MED, DS, Rician \textit{K}-factor, and link outage frequency for the Munich scene. Among the examined channel metrics, three pairs stand out with strong or considerable correlations: DS versus MED, MED versus Rician \textit{K}-factor, and PG versus link outage. It is important to note that, since many receiver locations can be considered to be at intermediate-distance from a given transmitter, the fitted ellipses to aggregate dispersion plots effectively show the behaviors at intermediate distances as well. 

The strong positive correlation between DS and MED implies that environments encouraging longer average arrival times naturally develop a more dispersed multipath profile. As it was previously noted, at intermediate distances small perturbations and geometric complexities can cause both the MED and DS to increase in variance. This suggests that sensitivity to height, position and material changes can produce regions where the timing of multipath arrivals becomes both later and more spread out.

The strong negative correlation between MED and Rician \textit{K}-factor points to a trade-off between temporal complexity and the dominance of a single strong path. When the average multipath arrival times increase, the channel tends to rely less on one standout component, which lowers the \textit{K}-factor. In previously discussed scenarios, at intermediate distances, multiple reflections emerge, making the channel more susceptible to perturbations. These conditions not only increase timing variability but also weaken the strength of any one dominant path, reflecting the interaction between geometry-driven complexity and the resulting temporal structure.

Finally, the considerable negative correlation between path gain and link outage shows the straightforward relationship between the two metrics as well as the importance of adequate received power for maintaining connectivity. Across varying distances, even as timing metrics become more sensitive, the probability of a link failing depends largely on whether sufficient power reaches the receiver. This shows maintaining adequate path gain emerges as a key factor that mitigates outage, highlighting how power-related metrics remain crucial despite increased temporal complexity in the channel.

\begin{table*}[h!]
\centering
\scriptsize
{\setlength{\tabcolsep}{3.2pt}
\caption{Standard Deviations of Channel Metrics for Different Scenes and Types of Perturbations. The average columns represent the square root of mean variance across all 10 transmitter locations. The min and max columns indicate the square root of mean variance for the transmitters exhibiting the minimum and maximum variations, respectively.}

\label{tab:std_dev_scene_type}
\begin{tabular}{|ll|ccc|ccc|ccc|ccc|}
\hline
Scene & Perturbation & $\sigma_{\text{PG,avg}}$(dB) & $\sigma_{\text{PG},\min}$ & $\sigma_{\text{PG},\max}$ & $\sigma_{\text{MED,avg}}$(ns) & $\sigma_{\text{MED},\min}$ & $\sigma_{\text{MED},\max}$ & $\sigma_{\text{DS,avg}}$(ns) & $\sigma_{\text{DS},\min}$ & $\sigma_{\text{DS},\max}$ & $\sigma_{\text{K,avg}}$(dB) & $\sigma_{\text{K},\min}$ & $\sigma_{\text{K},\max}$ \\
\hline
Munich & Material & 2.6 & 1.8 & 3.3 & 10.0 & 5.2 & 14.9 & 6.5 & 3.8 & 11.0 & 0.6 & 0.4 & 0.8 \\
Munich & Position & 3.1 & 2.4 & 3.9 & 144.4 & 105.6 & 217.5 & 60.0 & 33.1 & 92.0 & 6.6 & 4.9 & 8.3 \\
Munich & Height & 4.2 & 3.1 & 5.2 & 136.0 & 101.0 & 202.7 & 60.8 & 38.0 & 83.0 & 6.5 & 5.3 & 8.4 \\
Munich & Height and Position & 4.2 & 3.0 & 5.1 & 165.3 & 121.6 & 232.0 & 67.4 & 38.3 & 102.2 & 7.9 & 5.6 & 9.8 \\
Munich & Material, Height, Position & 4.2 & 3.0 & 5.3 & 166.7 & 117.0 & 239.8 & 67.4 & 37.2 & 105.6 & 7.8 & 5.5 & 9.7 \\
\hline
Etoile & Material & 2.0 & 1.5 & 2.7 & 13.4 & 2.1 & 15.5 & 9.9 & 1.4 & 11.9 & 0.6 & 0.3 & 1.0 \\
Etoile & Position & 2.6 & 1.7 & 3.3 & 163.1 & 28.0 & 204.0 & 75.4 & 11.7 & 90.1 & 5.1 & 3.2 & 6.0 \\
Etoile & Height & 4.0 & 3.1 & 6.1 & 143.9 & 21.9 & 182.5 & 76.8 & 15.5 & 99.5 & 5.6 & 4.7 & 7.0 \\
Etoile & Height and Position & 4.0 & 3.2 & 5.9 & 182.3 & 28.1 & 228.1 & 87.0 & 12.1 & 104.9 & 6.7 & 6.1 & 7.6 \\
Etoile & Material, Height, Position & 4.0 & 3.0 & 6.2 & 181.2 & 32.4 & 229.4 & 86.0 & 9.9 & 106.0 & 6.7 & 6.1 & 7.8 \\
\hline
\end{tabular}
}
\end{table*}

\subsection{Comparing Different Perturbation Types and Scenes}

Figures \ref{fig:broken_links_munich} and \ref{fig:broken_links_etoile} show that adjustments to building heights can have a larger effect on completely severing links, as compared to similarly scaled perturbations in building positions or material properties. One likely reason is that vertical modifications directly influence whether a LOS or key reflected path remains available: even a small elevation in building height can block a critical path segment, causing a direct communication failure. In contrast, horizontal shifts or slight variations in material properties might alter path timing and power levels but often do not eliminate the fundamental geometric feasibility of a link. Consequently, changes in height stand out as the most critical factor when evaluating link stability under uncertainty.

Table \ref{tab:std_dev_scene_type} compares the sensitivity of channel metrics across scenes and perturbation types: material, position, height, height and position combined, and all three combined. Comparing the two scenes and the different types of perturbations using this Table, it is evident that modifications to building geometry (height and/or position) have the most substantial impact on all channel metrics, while material perturbations alone tend to have a more modest influence. For instance, when looking at Munich, purely material-based changes result in an average PG standard deviation of around 2.6 dB, but introducing perturbations in building height alone pushes this figure to 4.2 dB. Similarly, for Etoile, material perturbations affect MED and DS only slightly (13.4 ns and 9.9 ns on average, respectively), but the inclusion of building height or position variations increases these values dramatically, pushing MED beyond 140 ns and DS beyond 75 ns on average. In both scenes, the Rician \textit{K}-factor exhibits a similar pattern, where geometric perturbations increase its standard deviation significantly more than adjustments in electromagnetic material properties.

Furthermore, while combining multiple types of perturbations—such as height and position, or material, height, and position—does not substantially increase the standard deviations of path gain beyond those seen with height or position variations alone, other metrics are more sensitive. Notably, MED and DS show increases that can approach or exceed the highest levels observed under single-type perturbations. This heightened sensitivity is particularly evident in the maximum values across different transmitter locations, where stacking multiple perturbations can push standard deviations to even greater extremes. Overall, these findings show that geometric fidelity is very important for achieving stable and reliable propagation predictions, while material accuracy, though less impactful for path gain variability, still plays a non-negligible role in shaping other critical channel statistics.

Examining the differences between position and height perturbations, it becomes clear that the magnitude of the perturbation matters: smaller positional uncertainties (0.4 meters) generally induce less variation in channel statistics than larger height uncertainties (1 meter). For both Munich and Etoile, height-only perturbations tend to produce higher standard deviations for all metrics compared to position-only changes. For instance, in Munich, position perturbations yield an average PG standard deviation of about 3.1 dB, whereas height perturbations increase the PG standard deviation to approximately 4.2 dB. A similar pattern holds for Etoile, where position perturbations result in an average PG standard deviation of 2.6 dB, while height perturbations reach 4.0 dB. This indicates that vertical adjustments to building structures, even if they seem minor (on the order of a meter), can more significantly alter the propagation environment than slightly shifting the horizontal positioning of the same structures. 

\subsection{Spatial Variability and Transmitter Location Dependence}

The minimum and maximum standard deviations reported in Table \ref{tab:std_dev_scene_type} reveal that transmitter placement also matters. Certain transmitter locations are relatively stable, showing minimal sensitivity to perturbations. Others are "hotspots" where small geometric or material modifications cause substantial variations in PG, MED, DS, and \textit{K}-factor. This emphasizes the importance of spatial context in predicting channel behavior and the need for careful transmitter siting and environmental modeling.


\section{Conclusion}

In conclusion, the geometry extraction process for ray-tracing propagation models is automated through a pipeline depicted in Figure \ref{fig:ge_pipeline}, which transforms user-specified region corner coordinates into detailed 3D scenes for electromagnetic simulations. By integrating various data sources—OpenStreetMap, Microsoft Global ML Building Footprints, and USGS elevation data—and automating the workflow via the Blender Python API, this pipeline reduces manual intervention and ensures consistent scene generation. 

In addition to streamlining geometry extraction, our findings show that accurate geometric modeling is crucial for reliable urban RF propagation predictions. While adjusting material properties can influence the temporal and power-related characteristics of the channel, even minor height or position changes often have a much stronger impact, especially in areas close to path-availability thresholds or in complex urban layouts with significant vertical scattering. Over intermediate transmitter-receiver distances, these height-driven changes can markedly increase variability in path gain, mean excess delay, delay spread, and link outage frequency.

Moreover, when the general material class of buildings is known, introducing a modest 10\% uncertainty in their relative permittivity and conductivity does not significantly change the resulting channel statistics. However, if the initial material parameters are largely unknown or mischaracterized, the resulting variability would likely be much greater. Therefore, it is important to have at least a baseline understanding of material properties to maintain stable and predictable propagation outcomes.

\bibliographystyle{IEEEtran}
\bibliography{biblio}

\end{document}